\documentclass[runningheads]{llncs}
\usepackage[T1]{fontenc}
\usepackage{graphicx}
\usepackage{xcolor}
\usepackage{amsmath,amssymb,amsfonts}
\usepackage{subcaption}
\usepackage[hidelinks]{hyperref}
\begin{document}

\captionsetup{font=small}
\captionsetup[sub]{font=small}
\title{A Comprehensive Analysis of Process Energy Consumption on Multi-Socket Systems with GPUs}
\titlerunning{Process Energy Consumption Analysis on Multi-Socket Systems with GPUs}

\author{Luis G. Le\'on-Vega\inst{1}\orcidID{0000-0002-3263-7853} \and Niccol\`o Tosato\inst{1,2}\orcidID{0009-0005-4996-9933} \and Stefano Cozzini\inst{2}\orcidID{0000-0001-6049-5242}}
\authorrunning{L. Le\'on-Vega et al.}
\institute{Università degli Studi di Trieste, Trieste, Italy \email{\{luisgerardo.leonvega,niccolo.tosato\}@phd.units.it} \\ \and AREA Science Park, Trieste, Italy \\
 \email{stefano.cozzini@areasciencepark.it}}

\maketitle              % typeset the header of the contribution
\begin{abstract}
Robustly estimating energy consumption in High-Performance Computing (HPC) is essential for assessing the energy footprint of modern workloads, particularly in fields such as Artificial Intelligence (AI) research, development, and deployment. The extensive use of supercomputers for AI training has heightened concerns about energy consumption and carbon emissions. Existing energy estimation tools often assume exclusive use of computing nodes, a premise that becomes problematic with the advent of supercomputers integrating microservices, as seen in initiatives like Acceleration as a Service (XaaS) and cloud computing. 

This work investigates the impact of executed instructions on overall power consumption, providing insights into the comprehensive behaviour of HPC systems. We introduce two novel mathematical models to estimate a process's energy consumption based on the total node energy, process usage, and a normalised vector of the probability distribution of instruction types for CPU and GPU processes. Our approach enables energy accounting for specific processes without the need for isolation.

Our models demonstrate high accuracy, predicting CPU power consumption with a mere 1.9\% error. For GPU predictions, the models achieve a central relative error of 9.7\%, showing a clear tendency to fit the test data accurately. These results pave the way for new tools to measure and account for energy consumption in shared supercomputing environments.

\keywords{Energy measurement \and, Supercomputers \and Microprocessors \and Multicore processing \and Load modeling \and High performance computing \and Cluster computing.}
\end{abstract}
\section{Introduction}

Server computers are widely used for cloud services, intensive computations, and network services, forming the foundation of modern computing. These computers often include powerful processing devices such as server-grade CPUs, large amounts of RAM, hardware accelerators, and cooling systems. With the arrival and rapid growth of Artificial Intelligence (AI) applications, server computers are in more demand, converting AI into one of the most demanding HPC workloads as the models become more complex and compute-hungry. It inspires a transition of HPC to Acceleration as a Service (XaaS) using microservices~\cite{hoefler2024xaas}, which implies sharing the computational resources amongst several services. This also raises an increasing concern about AI energy consumption and the carbon emissions produced to power these systems.

The current tools for measuring energy consumption in HPC are mostly limited to the node level. They are incapable of distinguishing the computation impacts, communication, and energy consumption distribution among the running processes. The absence of detailed granularity impedes our comprehension of the energy consumption characteristics of the applications and the identification of their most energy-intensive components. This research seeks to address this knowledge gap, thereby facilitating the development of more efficient and energy-conscious optimization strategies.

This work aims to comprehensively analyse the computing devices while executing a running process on a multi-socket computer. We then propose a novel energy model for estimating energy consumption while running a process of interest (PoI), focusing on CPU and GPU execution, pioneering in this research field. This model does not require execution isolation, making it ideal for shared computing nodes and the next generation of supercomputers based on XaaS.

\section{Background and Related Work}

Quantifying the energy consumption of computational systems is an open field that has been the focus of research across the wide spectrum of HPC, from embedded HPC to supercomputers. Most systems contain sensors that allow direct or indirect computation of the power drained by the various subsystems of the computer. This includes the processors, main memory, power supply units (PSU) and acceleration cards~\cite{psti,uprof,jetson}, covering measurements of the entire system at a course level.

A finer level of energy quantification involves measuring the energy consumption of the system’s running processes (or software tasks). Currently, there are still limitations in collecting this information. The hardware sensors cannot assign the power consumption to a running process, as this requires a software component that needs analysis.

Fine-grained energy consumption analyses typically focus on embedded systems, as they are simpler and often powered by batteries. These studies have analysed the impact of executing assembly instructions, such as load/storage~\cite{arm-load-store}, Very Long Instruction Words (VLIW)~\cite{xeon-phi,vliw-immediate-inst}, including some outdated GPUs~\cite{instruction-gpu}. Some proposals include power models to extract energy consumption based on the dispatched instructions, while others utilise heuristics at compile time~\cite{dev-compile}. These approaches are feasible given that the computing hardware (i.e. CPU, DSP or GPUs) consumes most of the power.

Particularly, from the GPU perspective, some studies evaluate the effect of memory transactions depending on the type of access~\cite{gpu-mem-access}, where some interesting insights have been found regarding the sequential and strided memory accesses. There are also studies analysing the kernel dispatching, suggesting that the multiple kernel execution also plays a role in how the energy is consumed~\cite{kernel-scheduling}. Others have benchmarked several workloads to get power profiles~\cite{benchmarks} and developed static analysis tools after a profile for predictions~\cite{static-analysys,gpu-prediction-model}, which are suitable for execution in isolation.

The scenario changes significantly for server-grade computers with increased complexity, where some fine-grained energy strategies are insufficient, unsuitable, or require further study. The study of the energy consumption in these computers must include cooling, storage, networking, memory and multi-socket CPU consumption. A survey from 2016 suggests that 50\% is used for cooling, 10\% for storage and 10\% for networking~\cite{survey-2016}. Another survey from 2020 reported that the CPU takes 32\% of the power consumption~\cite{survey-2020}, changing the scenario concerning the embedded system case, implying more components than only analysing the CPU's power.

The community has addressed estimating the energy consumed by server computers mainly by using models whose parameters include CPU utilization and frequency, memory and disk utilization, temperature, fan speed, and throughput (bandwidths)~\cite{survey-2020}. Others have been working on optimisation based on execution isolation and global utilisation for CPU and GPUs~\cite{micro-states}. Our key finding during our state-of-the-art study is that most models try to estimate energy at a system-wide level, and there has been little recent work on models that use the instructions executed by workloads as parameters, as found in embedded systems. Moreover, in the case of GPU energy accounting, models use similar ideas from the CPU, given its composition of multiple stream multiprocessors (or compute units)~\cite{survey-2016}. 

Additionally, the models proposed by related work are classified into different types, depending on the formulation. The most fundamental is the additive model, which describes the total power as the sum of all the loads~\cite{survey-2016,additive-model,power-addition}. The Baseline + Active (BA) models decompose the power load into base power (idle), active power (during computation), and a correction term, which can be interpreted as static power~\cite{survey-2016,ba-model}. Regression models are fitted using training data. The model can be linear, non-linear, or based on an existing power model~\cite{survey-2016}. However, all these models focus on general system-wide prediction rather than single-process estimation.

Therefore, there is an opportunity to contribute to the field of process-level energy consumption (fine-grained) and to use executed instructions as a parameter to enhance energy consumption estimation. This work will focus primarily on studying the energy impact of different types of instructions (scalar, vector, branching, and memory), the use of other computational resources, and how these can be used as parameters for energy consumption estimation of a single process by combining additive, BA and regression approaches.

\section{Process Power Consumption Analysis}

The instantaneous power consumption of a server computer can be modelled as an additive model, including the power consumption of the CPU, hardware accelerators (i.e. GPUs), RAM, storage, network interface cards (NIC), cooling (fans), and other electronic components~\cite{power-addition}, which involves the following power model at an instant $t$:
\begin{multline}
    P_{\text{system},t} = P_{\text{CPU,t}} + P_{\text{Accel,t}} + P_{\text{RAM,t}} + P_{\text{Disk,t}} + P_{\text{NIC,t}} + P_{\text{Cool,t}} + P_{\text{Aux,t}}
\end{multline}

All these power components can vary their values over time based on their utilisation and power domain status (on or off). For instance, CPU-intensive workloads will cause an increase in the $P_{\text{CPU,t}}$, as well for HW-accelerator workloads, which increase the $P_{\text{Accel,t}}$. Nevertheless, the increase starts from a base power value, which corresponds to the idle state of the device that matches the static power. In this case, the first proposal is to decompose each power component into its static and dynamic components, where the latter depends on the activity of the device:
\begin{equation}
    P_{\text{device},t} = P_{\text{dynamic},t} + P_{\text{static},t}
\end{equation}

This corresponds to a BA model~\cite{ba-model}, that can be even more aware of the status of the hardware, taking into consideration the existence of power domains that can be switched off for energy savings:
\begin{equation}
    P_{\text{device},t} = \big(P_{\text{dynamic},t} + P_{\text{static},t}\big)u + P_{\text{suspend},t}\big(1-u\big)
\end{equation}

\noindent
where $u = \{0,1\}$ is a status variable, that takes a $0$ value if the device if suspended or $1$ if it is on. For this work, we will assume that the device is always on ($u = 1$) and will revisit this assumption in future work.

Some workloads utilise more than one device at a time. For instance, a CPU-based matrix multiplication uses the CPU and the RAM (if all operands are loaded into memory). Therefore, we propose classifying the workloads based on their computational nature, i.e. CPU-based workloads, GPU-based workloads, storage operations and network communication. This is possible due to the conservation of energy, which states that the power supplied by a power source is equal to the sum of all the power loads. Moreover, it is possible to characterise any electrical circuit using the superposition principle, meaning that exercising parts of the circuit gives a characterisation of their loads~\cite{power-addition}.

This section will be divided into the computational natures (CPU and GPU), analysing each component and using actual measurements to define the behaviours of the loads, using a set of benchmarks that can exercise the several parts of the computer. For the CPU, we are specifically using the hardware from Table~\ref{tbl:hardware-epyc}, and for the GPU, the hardware from Table~\ref{tbl:hardware-gpu}. Network and disk activity are not within the scope of this work and will be addressed in future analysis.

\begin{table}[t]
\scriptsize
\centering
\caption{PowerEdge R6525 Hardware Configuration}
\begin{tabular}{|l|c|}
\hline
\textbf{Component} & \textbf{Description}                          \\[3pt] \hline
CPU & 2 X AMD EPYC 7H12 64 cores                          \\[3pt] \hline
RAM     & 16 X 32GB of 3200 MT/s DDR4 RAM ECC                            \\[3pt] \hline
Disk & 2 X 480GB SSD -  RAID 0                          \\[3pt] \hline
NIC               & MT28908 Family [ConnectX-6] 100Gb (Infiniband)                     \\[3pt] \hline
Cooling                & 16 x Dell High performance (Silver grade) fan                             \\[3pt] \hline
PSU                &  2 x 1400 W redundant PSU                           \\[3pt] \hline

\end{tabular}
\label{tbl:hardware-epyc}
\end{table}

\begin{table}[t]
\scriptsize
\centering
\caption{PowerEdge R740xd Hardware Configuration }
\begin{tabular}{|l|c|}
\hline
\textbf{Component} & \textbf{Description}                          \\[3pt] \hline
CPU & 2 X Intel(R) Xeon(R) Gold 6226 CPU @ 2.70GHz 12 cores                          \\[3pt] \hline
RAM     & 8 X 32GB of 2933 MT/s DDR4 RAM ECC                            \\[3pt] \hline
GPU     & 2 X NVIDIA Tesla V100 PCIe 32GB                   \\[3pt] \hline
Disk & 2 X 480GB SSD -  RAID 0                          \\[3pt] \hline
NIC               & MT28908 Family [ConnectX-6] 100Gb (Infiniband)                     \\[3pt] \hline
Cooling                & 6 x Performance Fans for R740/740XD                             \\[3pt] \hline
PSU                &  2 x 1600 W redundant PSU                           \\[3pt] \hline

\end{tabular}
\label{tbl:hardware-gpu}
\end{table}

\subsection{Energy Model for CPU workloads}

For this work, the first assumption is that we can obtain metrics of the total CPU consumption and the power delivered by the PSU. For simplicity, we will exclude other computing components like GPUs during this analysis and use the superposition principle explained earlier. In this context, we can rearrange the model as follows:
\begin{equation}
    P_{\text{system},t} \approx P_{\text{PSU},t} = P_{\text{CPU,t}} + P_{\text{Other,t}}
\end{equation}

\noindent
where $P_{\text{Other,t}}$ encapsulates the consumption of all the hardware except for the CPU. The CPU consumption involves the addition of non-processing components, such as memory and peripheral controllers, plus the consumption of each core:
\begin{equation}
    P_{\text{CPU,t}}(\mathbf{f}, \mathbf{T}, \mathbf{w}) = 
    \sum_{i=1}^{N_c} P_{\text{core},t}^{i}(f^i, T^i, w^i) + P_{\text{CPU,other},t}(\mathbf{f}, \mathbf{T})
    \label{eqn:total-cpu}
\end{equation}

\noindent
where $P_{\text{core},t}^{i}$ shows the power consumption of the $i$-th core and $P_{\text{CPU,other}}$ represents the power consumed by the other parts that integrate the processor~\cite{power-base-mem-access}. In this model, we consider that the frequency ($f$), temperature ($T$), and system workload ($w$) can be different for each core, represented as vectors ($\mathbf{f}, \mathbf{T}, \mathbf{w}$, respectively). Assuming that $P_{\text{CPU,other}}$ is not part of any core and does not depend on the $\mathbf{w}$ and most of the power is consumed by the cores, we can focus on these latter components, which can be approximated by a BA submodel:
\begin{equation}
    P_{\text{core},t}^{i}(f, T, w) = P_{\text{static},t}^{i}(f, T) +  P_{\text{dynamic},t}^{i}(f, T, w)
    \label{eqn:total-core}
\end{equation}

\noindent
where $P_{\text{static},t}^{i}$ represents the static power consumed by the core and does not depend on the system workload, and $P_{\text{dynamic},t}^{i}$ the dynamic power~\cite{power-usage-var,power-core-decomp}. Up to this point, we pursue the dynamic power, which varies depending on the frequency, operation type, workload, and temperature. One of the most comprehensive dynamic power models is given by $P_{\text{dynamic},t}^{i} = y_i \alpha_i f_i^{\beta_i}$, where $y_i$ is the core state, $\alpha_i$ and $\beta_i$ are system-specific parameters associated to the voltage, switching activity (implicitly the workload) and capacitance of the transistors, and $f_i$ is the core frequency~\cite{power-core-decomp}.

To handle the CPU usage, some works estimate the average full-load dynamic power $P_{\text{max(dynamic)},t}^{i}$ and multiply it by the usage, such that $P_{\text{dynamic},t}^{i} = w^i P_{\text{max(dynamic)},t}^{i}$, where $w^i$ is the workload measured as core utilisation from 0 to 1~\cite{power-usage-var}. Nevertheless, this approximation is inaccurate since the core utilisation is often represented by the portion of the time the CPU is active in a given period, and the type of instruction executed by the CPU is not considered. The operating system (OS) may report 100\% CPU utilisation on scalar operations and 100\% on vector operations, yet energy consumption varies between these numbers.

The second assumption in our power model is that the power per core cannot be quantified, and the CPU runs at the base clock at an average temperature with low variance. This is a limitation of some systems that do not have instruments per core, like ARM-based systems or x86 which provide metrics based on heuristics. It simplifies the analysis for further steps and the scope of this work. This assumption will be left for future work. In this case, we need to refer back to equation (\ref{eqn:total-cpu}), such that
\begin{multline}
    P_{\text{CPU,t}}(\bar{f}, \bar{T}, \hat{w}) =
    P_{\text{dynamic},t}(\bar{f}, \bar{T}, \hat{w}) + P_{\text{static},t}^{i}(\bar{f}, \bar{T}) +
    P_{\text{CPU,other},t}(\bar{f}, \bar{T})
\end{multline}

\noindent
where equation (\ref{eqn:total-core}) is reformulated to rewrite the sum of the dynamic power per core as the dynamic power of the entire CPU, which now depends on the mean frequency ($\bar{f}$) and temperature $\bar{T}$ and the total CPU workload $\hat{w}$. Likewise, we can encapsulate the sum of the static power per core to be the $P_{\text{CPU,other},t}(f^i, T^i)$, assuming that the entire CPU will be always on, therefore
\begin{equation}
    P_{\text{CPU,t}}(\bar{f}, \bar{T}, \hat{w}) = 
    P_{\text{dynamic},t}(\bar{f}, \bar{T}, \hat{w}) + P_{\text{CPU,other},t}(\bar{f}, \bar{T})
    \label{eqn:cpu-dynamic}
\end{equation}

Building on the model from equation (\ref{eqn:cpu-dynamic}), we aim to define the dynamic power of the CPU ($P_{\text{dynamic},t}$). We propose introducing a more comprehensive utilisation parameter based on the types of instructions executed in a time window, such as the scalar/vector arithmetic, logic, memory and branching, and distributing the utilisation of the CPU and power consumption according to the impact of the instruction types. This modifies (\ref{eqn:cpu-dynamic}) to
\begin{equation}
    P_{\text{CPU,t}}(\bar{f}, \bar{T}, \hat{w}, \mathbf{h}) = 
    P_{\text{dynamic},t}(\bar{f}, \bar{T}, \hat{w}, \mathbf{h}) + P_{\text{CPU,other},t}(\bar{f}, \bar{T})
\end{equation}

\noindent where the dynamic power is
\begin{equation}
    P_{\text{dynamic},t}(f, T, w, \mathbf{h}) = l(w, \mathbf{h}, f, T) \theta(f, T) \bar{P}_{\text{max(dynamic)}}(f, T)
\end{equation}

\noindent
$l(w, \mathbf{h}, f, T)$ is a function that quantifies the impact of each instruction given a histogram vector $\mathbf{h}$ in a time window, $\theta(f, T)$ is a non-linear function that depends on the CPU intrinsic parameters, like electrical impedance, $\bar{P}_{\text{max(dynamic)}}$ the mean power when the CPU is loaded to its maximum capacity and $w$ is the CPU utilisation reported by the OS, whose values can oscillate between $0$ and $N_c$, where $w = N_c$ means that all CPU cores are being used 100\%.

For the assumptions made so far, the CPU's frequency is fixed to a single value, and the temperature is almost constant by setting the fans to the maximum speed during the experiment. It simplifies the problem of finding the function $l$ that describes the impact of the instruction and the process utilisation. Besides, we are grouping the instructions into scalar and vector families and arithmetic, logic, memory and branch types, where each family has all instruction types. Therefore, the CPU power can be defined as
\begin{equation}
    P_{\text{CPU},t}(w, \mathbf{h}) = \hat{\theta} l(w, \mathbf{h}) \bar{P}_{\text{max(dynamic)}}  + \hat{P}_{\text{CPU,other},t}
    \label{eqn:cpu-freqless-templess}
\end{equation}

\noindent
where $\hat{\theta}$ is an estimation parameter that represents the CPU intrinsic parameters and their affectation by the operating frequency and temperature, and $\hat{P}_{\text{CPU,other},t}$ is an estimated parameter that represents the static and base power of the CPU. The equation (\ref{eqn:cpu-freqless-templess}) can be decomposed to apply a linear regression with a transformation on $w$ and $\mathbf{h}$ that can describe possible non-linearities:
\begin{equation}
    P_{\text{CPU},t}(w, \mathbf{h}) = \sum_{k=1}^{S} \gamma_{k} \sigma(h_k, w) + \hat{P}_{\text{CPU,other},t}
    \label{eqn:regression-model}
\end{equation}

\noindent
where $\gamma_k$ are the coefficients to determine through linear regression, compressing $\hat{\theta}$ and $\bar{P}_{\text{max(dynamic)}}$, $S$ is the number of instruction types and $\sigma$ is a function that maps each instruction type and the process utilisation. The linear regression also determines the intercept $\hat{P}_{\text{CPU,other},t}$, corresponding to the processor's idle consumption.

Equation (\ref{eqn:regression-model}) must adhere to energy conservation principles. It implies that the total dynamic power consumed by $N_p$ processes is the result of the sum of all the dynamic power consumed by each process $p$:
\begin{equation}
    P_{\text{CPU},t} = \sum_{p=1}^{N_p}\sum_{k=1}^{S} \gamma_{k} \sigma(h_k^{p}, w_p) + \hat{P}_{\text{CPU,other},t}
    \label{eqn:regression-model-process}
\end{equation}

\noindent
where $N_p$ is the number of processes, $h_k^p$ is the probability of the instruction type $k$ in the process $p$ and $w_p$ is the CPU utilisation of the process $p$. Therefore, the model from equation (\ref{eqn:regression-model-process}) assumes that we have measurements of the power from either the PSU or the CPU; the frequency and the temperature are fixed, with low variance and the same for all cores, the power domain is always on, and the instructions can be collected for the processes using an OS tool. The model can be extended to remove the assumptions regarding frequency, temperature, and power domain.

\begin{figure}[!h]
    \begin{subfigure}[b]{0.48\columnwidth}
        \centering
        \includegraphics[width=0.99\columnwidth]{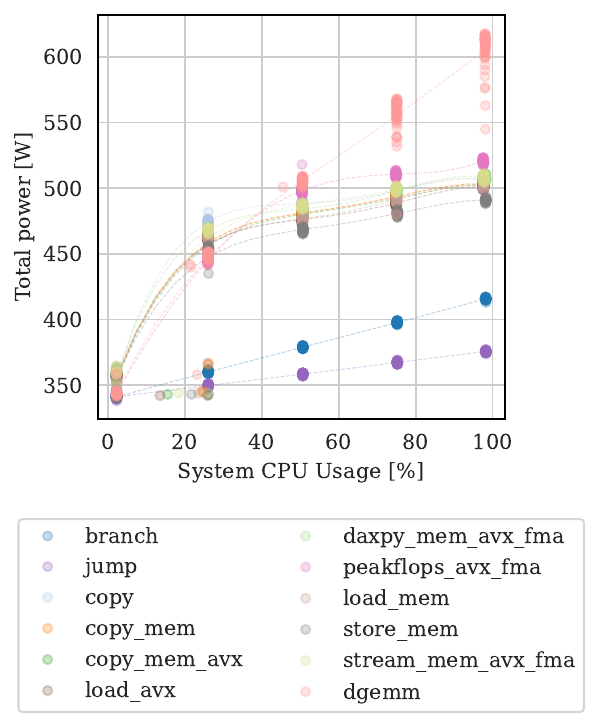}
        \caption{CPU Power Consumption}
        \label{fig:power-consumption-system-usage}
    \end{subfigure}\hfill
    \begin{subfigure}[b]{0.52\columnwidth}
        \centering
        \includegraphics[width=0.97\columnwidth]{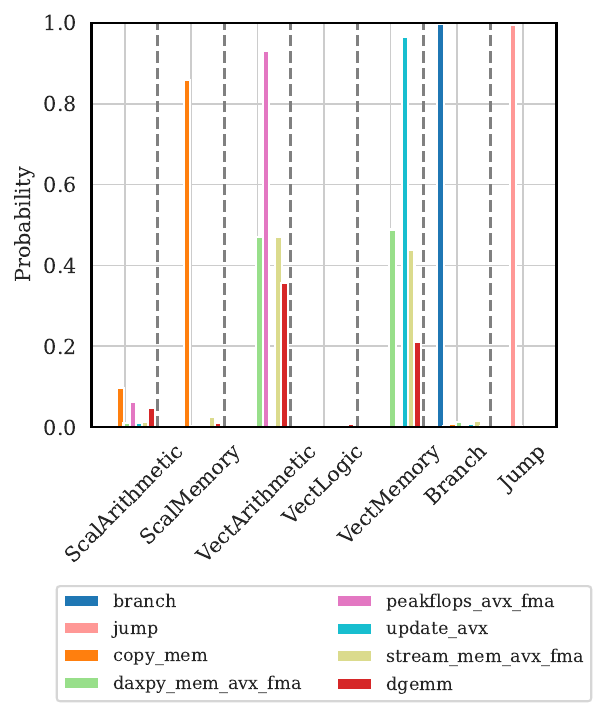}
        \caption{Histogram of Operation Types}
        \label{fig:histogram}
    \end{subfigure}
    \caption{Experiments behaviour when scaling in degree of parallelism (System CPU Usage) and the instructions executed. On the left, the socket power consumption is reported as the usage scales, whereas on the right, the histogram of the instructions illustrates the experiment's nature.}
    \label{fig:cpu-experiments}
    \vspace{-5mm}
\end{figure}

Fig.~\ref{fig:cpu-experiments} depicts a series of experiments that perform computations using scalar, vector and memory operations. Fig.~\ref{fig:power-consumption-system-usage} shows how the aggregated CPU socket consumption increases as the system CPU utilisation scales, illustrating a logarithmic behaviour in most experiments except in the branch and jump experiments, where the behaviour is linear. Moreover, each experiment has a different socket consumption, highlighting the role of the instructions, illustrated in Fig.~\ref{fig:histogram}, on the overall power consumption. Fig.~\ref{fig:histogram} serves as a reference to highlight that each experiment has a different probability of launching certain types of instructions. The data was collected using EfiMon~\cite{efimon} and the machine from Table~\ref{tbl:hardware-epyc}.

We propose to model the impact of the instruction histograms and the utilisation as a hybrid linear-logarithmic combination, where the impact of the instruction is also conditioned by the degree of parallelism, given implicitly by the system usage, which denotes how many CPU cores are used during the execution. Therefore, the $\sigma$ functions that summarises the impact of $w$ and $\mathbf{h}$ is:
\begin{equation}
    \sigma(h_k, w_p)_1 = h^p_k\ln(w_p + 1)
    \label{eqn:sigma-def-1}
\end{equation}
\begin{equation}
    \sigma(h_k, w_p)_2 = h^p_k w_p
    \label{eqn:sigma-def-2}
\end{equation}

\noindent
where $k$ and $p$ are the indices for the instruction type (classified into scalar and vector types, and memory, branching, arithmetic and logic families) and the process index, respectively. The $\sigma(h_k, w_p)_1$ is used by the memory instructions (scalar and vector) and the vector arithmetic. The $\sigma(h_k, w_p)_2$ is used for the rest of instruction types (scalar arithmetic, logic, branch and jumps).

On the other hand, considering our previous statement that, in a computing execution, more than one device is involved, the equation can be generalised to include other peripherals, such as caches, memories, and controllers, leading to:
\begin{equation}
    P_{\text{PSU},t} = \sum_{p=1}^{N_p}\sum_{k=1}^{S} \gamma_{k}  \sigma(h_k^{p}, w_p) + \hat{P}_{\text{other},t}
    \label{eqn:regression-model-process-complete-psu}
\end{equation}

\noindent
where the model includes the PSU and other peripherals, including the CPU, RAM, and other electronics with a passive consumption in $\hat{P}_{\text{other},t}$. In order to perform a regression of the model, we can stick to the isolation principle, executing a process at a time, thanks to the superposition principle, allowing the removal of the sum over $p \in N_p$. Finally, in order to extract the process energy consumption, we can use the calculated regression model and the dynamic part:
\begin{equation}
    P_{\text{p},t} = \sum_{k=1}^{S} \gamma_{k}  \sigma(h_k^{p}, w_p)
    \label{eqn:regression-model-process-complete-pid}
\end{equation}

\subsection{Energy Model for GPU workloads}

Graphics Processing Units (GPUs) can be seen as a particular type of CPU architecture with massive execution units. According to NVIDIA naming conventions, a GPU is composed mainly of Stream Multiprocessors (SM), a DRAM memory bank, and caches. The SMs can be likened to simplified CPU processing cores with in-order execution, having their own scheduler, caches, and many execution units, usually a multiple of the warp size (32). Unlike CPUs, each SM launches many threads that execute the same instruction simultaneously. This is the opposite of how CPUs assign threads for execution; CPUs assign threads to each core, whereas GPUs collect instructions from threads to execute them all simultaneously~\cite{nvidiacuda,cuda}. Recent architectures have optimised this process to avoid underutilisation~\cite{cuda}.

To the best of our knowledge, non-invasive fine-grained GPU energy measurement has not been explored using online analysis. However, a tool within the CUDA toolchain called NVML~\cite{nvml} captures information about GPU energy consumption, temperatures, utilization, and clock speeds. Furthermore, with the assumption stated above regarding the architecture, we can proceed with an analysis similar to the one performed for the CPU. Equation (\ref{eqn:regression-model}) shows the power consumed by the CPU. We can use this equation as a reference, considering that the GPUs are often expansion cards that communicate with the host systems through PCIe. In the case of NVIDIA GPUs, they have power meters to account for the energy consumption of the entire GPU module~\cite{nvml}. Therefore, the power consumption of the GPU card can be expressed as:
\begin{equation}
    P_{\text{GPU},t}(w, \mathbf{h}) = \sum_{k=1}^{S} \gamma_{k} \sigma(h_k, w) + \hat{P}_{\text{GPU,other},t}
    \label{eqn:regression-model-gpu}
\end{equation}

\noindent
where $\mathbf{h}$ is the instruction histogram obtained from the acceleration kernel object (GPU assembly code), $\sigma$ is a function that summarises the behaviour of each instruction type $h_k$ and the GPU compute usage $w_p$ linked to the process $p$ to the power consumption, $\gamma_k$ represents the impact of the instruction on the power consumption and $\hat{P}_{\text{GPU,other},t}$ represents the power consumption from non-computational peripherals inside of the GPU card.

\begin{figure}[!h]
    \vspace{-5mm}
    \begin{subfigure}[b]{0.45\columnwidth}
        \centering
        \includegraphics[width=0.95\columnwidth]{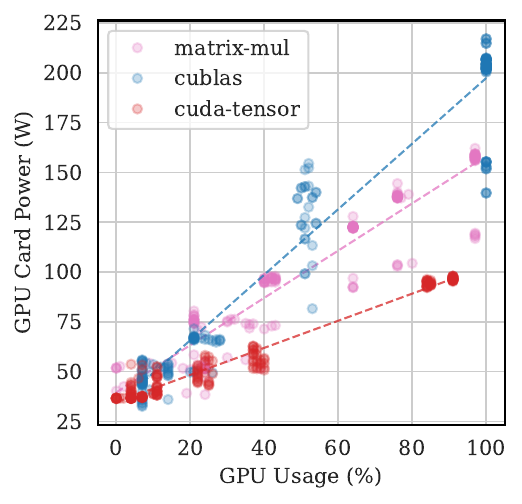}
        \vspace{7mm}
        \caption{GPU Power Consumption}
        \label{fig:gpu-power-consumption-system-usage}
    \end{subfigure}\hfill
    \begin{subfigure}[b]{0.54\columnwidth}
        \centering
        \includegraphics[width=0.95\columnwidth]{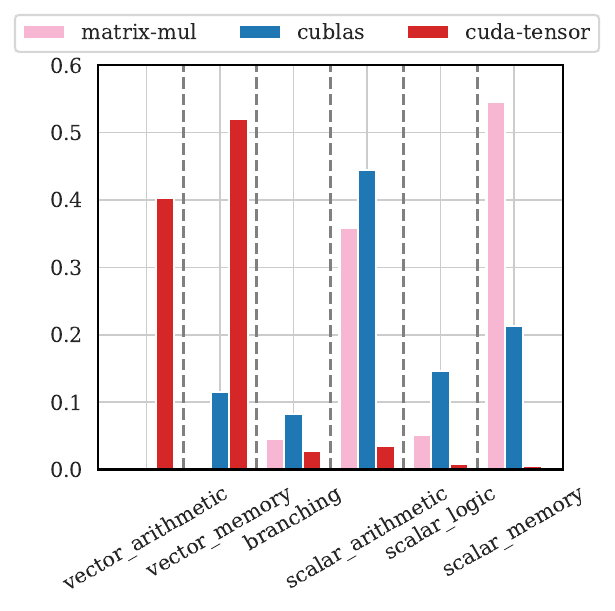}
        \caption{Histogram of Operation Types}
        \label{fig:gpu-histogram}
    \end{subfigure}
    \caption{Experiments behaviour when scaling the GPU occupation and the instructions. On the left, the GPU card power consumption is reported as the usage scales, whereas on the right, the histogram of the instructions illustrates the experiment's nature.}
    \label{fig:gpu-experiments}
    \vspace{-5mm}
\end{figure}

Similarly to the CPU case, this work performs a series of experiments that stress different GPU instructions. For this case, different implementations of a $4096 \times 4096 \times 4096$ matrix multiplication were explored, with emphasis on memory instructions, vector instructions (Tensor Cores from NVIDIA~\cite{v100}) and an optimised algorithm in CUBLAS~\cite{cuda}. Fig.~\ref{fig:gpu-experiments} shows the experiments' behaviour with respect to their power consumption and the histogram of the instructions executed by each implementation. While performing the same matrix multiplication operation workload, it is possible to observe that each experiment uses different instructions to perform the same task and consumes power differently. The CUBLAS implementation has more scalar arithmetic and is the most power-consuming. On the other hand, the Matrix Multiplication is a hand-crafted kernel that stands as a middle point in power consumption but has the most memory transactions. The implementation based on CUDA Tensor Cores is the most energy efficient, as it uses the Tensor Cores (a special execution unit for SIMD operations) and has the least power consumption.

To determine the nature of $\sigma$, we can perform a variable analysis to see how the product between the probability of the instruction $k$ and the GPU usage $w$ is related to the power consumption. From Fig.~\ref{fig:gpu-power-consumption-system-usage}, the variables involved in the analysis, such as the utilisation and the instruction affectation, exhibit linear behaviour, suggesting that the product between the instruction and the GPU usage has a proportional effect on the power consumption. Therefore, we can assume that $\sigma$ for the GPU is:
\begin{equation}
    \sigma \big(h_k, w_p\big) = h^p_k w_p
    \label{eqn:linear-gpu}
\end{equation}

This leads to the following power consumption model for a granular GPU utilisation:
\begin{equation}
    P_{\text{GPU},t} = \sum_{p=1}^{N_p}\sum_{k=1}^{S} \gamma_{k} h_k^{p}w_p + \hat{P}_{\text{GPU,other},t}
    \label{eqn:regression-model-process-complete-gpu}
\end{equation}

\noindent
where the power consumption for a single process running on the GPU can be expressed as:
\begin{equation}
    P_{\text{p},t} = \sum_{k=1}^{S} \gamma_{k} h_k^{p}w_p
    \label{eqn:regression-model-process-complete-pid-gpu}
\end{equation}

\noindent
which is useful for a linear regression study to determine the parameters $\gamma_k$ and the intercept $\hat{P}_{\text{GPU,other},t}$ to complete the model. A relevant side note to highlight is that, if the power measurement during the linear regression is accounted at the card level, the $\gamma_k$ includes the impact of executing a process on the GPU compute units and the video RAM included in the GPU.

%NVML provides: clock speed, fan speed, power consumption, process consumption and memory consumption

\section{Model Experimental Results}

This section explores the fitting of the linear regression models for the CPU and the GPU, represented by equations (\ref{eqn:regression-model-process-complete-psu}) and (\ref{eqn:regression-model-process-complete-gpu}), respectively. For this purpose, we use the hardware presented in Table~\ref{tbl:hardware-epyc} for the CPU experiment and Table~\ref{tbl:hardware-gpu} for the GPU. Moreover, we use a dataset created from the latter hardware, taking 50 samples per experiment run (number of threads).

\begin{figure}[!h]
    \begin{subfigure}[b]{0.54\columnwidth}
        \centering
        \includegraphics[width=0.95\columnwidth]{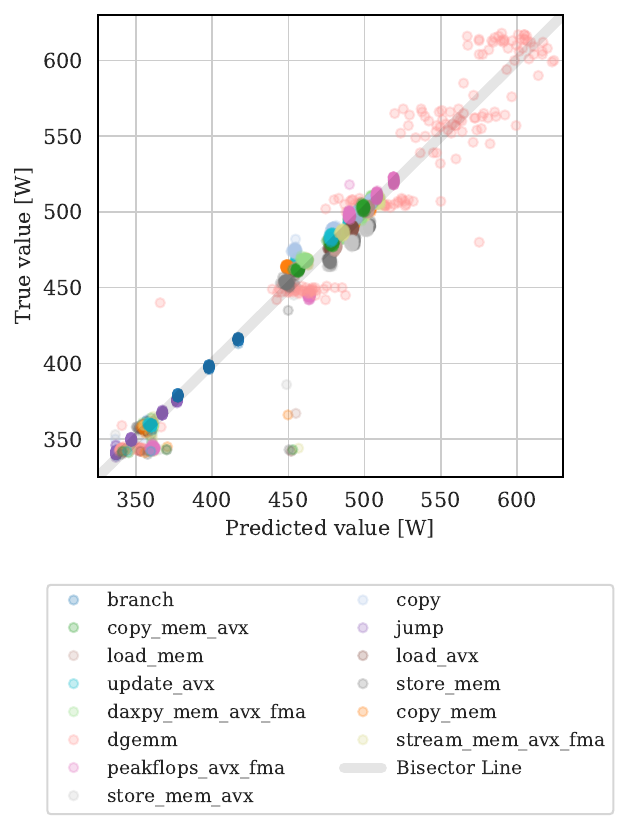}
        \caption{CPU Model}
        \label{fig:cpu-model}
    \end{subfigure}\hfill
    \begin{subfigure}[b]{0.46\columnwidth}
        \centering
        \includegraphics[width=0.95\columnwidth]{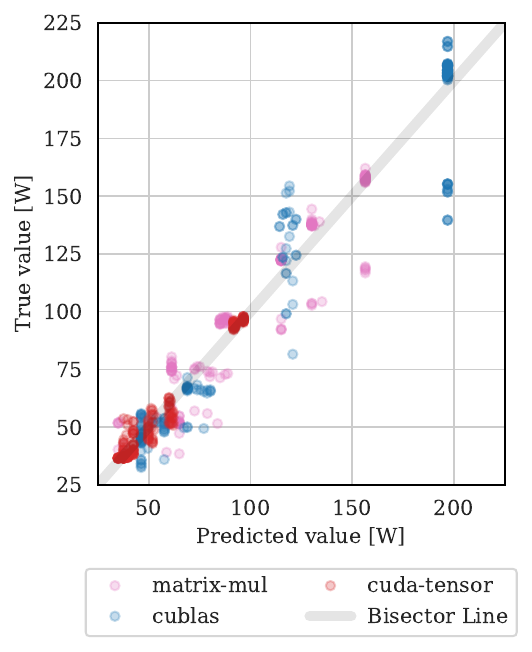}
        \caption{GPU Model}
        \vspace{19mm}
        \label{fig:gpu-model}
    \end{subfigure}
    \caption{CPU and GPU model performance in power consumption prediction. On the left, the CPU model is fitted using non-negative linear regression. On the right, the GPU model is fitted using least-squares linear regression.}
    \label{fig:fit}
    \vspace{-8mm}
\end{figure}

Fig.~\ref{fig:fit} shows the fitted models for the CPU and GPU. The CPU regression model (Fig.~\ref{fig:cpu-model}) has a root mean square error (RMSE) of 9.69 Watts for the dataset, showing outstanding prediction performance for the whole system when relating the instructions and the CPU usage. The relative error of this RMSE value with respect to the midpoint of the dataset (500 W) corresponds to 1.93\%. This leads to an opportunity to contribute to fine-grained energy prediction using runtime available data, enabling measurements of the energy footprint of the computations on the CPU side.

In contrast, the GPU model has an RMSE of 12.3 Watts, around 1.27 times less effective than the CPU model. This corresponds to a 9.7\% of relative error with respect to the central point of the dataset. Besides, Fig.~\ref{fig:gpu-model} shows that most of the points concentrate on the bisection line, which represents the optimal line of the model. This suggests that the outliers play a role in affecting the overall model performance. The outliers are generated due to the runtime metrics, influenced mainly by the sampling time of the kernel execution, the power accounting tool and the profiler used to take the measurements.

\begin{table}[t]
\centering
\caption{Parameters Estimated using Least Squares Linear Regression for models using CPU Power Model from equation (\ref{eqn:regression-model-process-complete-psu}) and GPU model from (\ref{eqn:regression-model-process-complete-gpu}) as prediction reference.}
\vspace{3mm}
\scriptsize
\begin{tabular}{|l|c|c|}
\hline
\textbf{Parameter Name}                     & \textbf{CPU Power Model}   & \textbf{GPU Power Model}    \\[3pt] \hline
Intercept ($\hat{P}_{\text{static},t})$     & 336.5031                   & 34.9818     \\[3pt] \hline
Weight of Scalar Arithmetic ($\gamma_{sa}$) & 0.6717                     & 276.1728    \\[3pt] \hline
Weight of Scalar Memory ($\gamma_{sm}$)     & 35.6589                    & 33.0339     \\[3pt] \hline
Weight of Scalar Logic ($\gamma_{sl}$)      & 0.00000                    & 108.412     \\[3pt] \hline
Weight of Vector Arithmetic ($\gamma_{va}$) & 38.6822                    & 4.9488      \\[3pt] \hline
Weight of Vector Memory ($\gamma_{vm}$)     & 35.3435                    & 102.3084    \\[3pt] \hline
Weight of Vector Logic ($\gamma_{vl}$)      & 154.5258                   & 0.0000      \\[3pt] \hline
Weight of Branch ($\gamma_{vl}$)            & 0.6459                  & 0.00000     \\[3pt] \hline
Weight of Jumps ($\gamma_{vl}$)             & 0.3239                  & 0.00000     \\[3pt] \hline
\end{tabular}
\label{tbl:parameters}
\end{table}

Table~\ref{tbl:parameters} shows the parameters obtained after the Least Squares Linear Regression for the models described by equations (\ref{eqn:regression-model-process-complete-psu}) and (\ref{eqn:regression-model-process-complete-gpu}). The model's intercept gives the system's static power, which in idle is 336 Watts, whereas, for the GPU card, it is 34.98 Watts.

In terms of relevance, the CPU model places more importance on vector logic operations, followed by vector arithmetic, which are linear factors. It is also impacted by memory, which is represented as the product of the probability and the logarithm of usage. This is consistent with the fact that vector instructions are the most power-consuming in CPUs, followed by the memory transactions, which involve changes in the DRAM and various queries along the memory hierarchy~\cite{intel_avx_whitepaper,intel_xeon_specification}.

In the GPU model, the most relevant instructions are the scalar, involving the CUDA cores and memory transactions. The vector arithmetic seems less relevant given that the Tensor Cores are more optimised than the CUDA cores~\cite{v100}. It is also consistent with the observations from Fig.~\ref{fig:gpu-power-consumption-system-usage}, suggesting the hardware optimisations in the GPU under study. Likewise, the vector memory loads are more relevant than the scalar version because they can spend more cycles filling different registers. 

\section{Conclusions and Future Work}

This work has presented a novel framework for instruction-driven process energy consumption analysis in multi-socket computers with GPU accelerators in shared computational environments. We developed two mathematical models (one for CPU and another for GPU) to estimate the power consumption of individual processes. These models are based on process usage, instruction histograms, and power meter data, making them highly adaptable for real-world HPC applications. The CPU model was trained on a dual-socket compute node using a set of likwid benchmarks, while the GPU model was trained on an NVIDIA Tesla V100 using three different matrix multiplication implementations. 

Our experimental results showed outstanding performance, with the CPU model achieving a relative error of 1.93\% and the GPU model achieving a relative error of 9.7\%. These results highlight the effectiveness of our models and their potential to significantly advance energy accounting in HPC systems. By enabling measurement of the energy consumption of individual processes, our models isolate the effects of different processes running on the same system, paving the way for more efficient and energy-conscious optimisation strategies.

In future work, we plan to integrate the frequency and fan speed as variables in the model, making it more adequate in real scenarios these variables fluctuate due to runtime energy optimisations. Our approach holds great promise for developing new tools and methodologies that enhance energy efficiency in shared supercomputing environments.

\begin{credits}
\subsubsection{\ackname} Results achieved with the funding obtained under Axis IV of the PON Research and Innovation 2014-2020 "Education and research for recovery - REACT-EU". Experimental results correspond to the ORFEO Supercomputer at AREA Science Park. The project was partly funded by eXact Lab s.r.l.

\subsubsection{\discintname}
The authors have no competing interests to declare that are relevant to the content of this article.
\end{credits}
\bibliographystyle{splncs04}
\bibliography{main}
\end{document}